\documentclass[%
 reprint,
 superscriptaddress,
 amsmath,amssymb,
 aps,
]{revtex4-2}

\usepackage{graphicx}
\usepackage{dcolumn}% Align table columns on decimal point
\usepackage{bm}% bold math
\begin{document}

\preprint{APS/123-QED}

\title{Evidence for positive long- and short-term effects of vaccinations against COVID-19 in wearable sensor metrics --- Insights from the German Corona Data Donation Project}

\author{Marc Wiedermann}\email{marcw@physik.hu-berlin.de}
\author{Annika H. Rose}
\author{Benjamin F. Maier}
\author{Jakob J. Kolb}
\author{David Hinrichs}
\affiliation{Institute for Theoretical Biology und Integrated Research Institute for the Life-Sciences, Humboldt University of Berlin, Philippstr. 13, 10115 Berlin, Germany}
\affiliation{Robert Koch Institute, Nordufer 20, 13353 Berlin, Germany}

\author{Dirk Brockmann}
\affiliation{Institute for Theoretical Biology und Integrated Research Institute for the Life-Sciences, Humboldt University of Berlin, Philippstr. 13, 10115 Berlin, Germany}

\date{\today}

\begin{abstract}
Vaccines are among the most powerful tools used to combat the COVID-19 pandemic. They are highly effective against infection and substantially reduce the risk of severe disease, hospitalization, ICU admission, and death. However, their potential for attenuating long-term effects of a SARS-CoV-2 infection, commonly denoted as Long COVID, remains elusive and is still subject of debate. Such long-term effects can be effectively monitored at the individual level by analyzing physiological data collected by consumer-grade wearable sensors. Here, we investigate changes in resting heart rate, daily physical activity, and sleep duration in response to a SARS-CoV-2 infection stratified by vaccination status. Data was collected over a period of two years in the context of the German Corona Data Donation Project with currently around 190,000 monthly active donors. Compared to their unvaccinated counterparts, we find that vaccinated individuals on average experience smaller changes in their vital data that also return to normal levels more quickly. Likewise, extreme changes in vitals during the acute phase of the disease occur less frequently in vaccinated individuals. Our results solidify evidence that vaccines can mitigate long-term detrimental effects of SARS-CoV-2 infections both in terms of duration and magnitude. Furthermore, they demonstrate the value of large scale, high-resolution wearable sensor data in public health research.
\end{abstract}

\maketitle

\section{Introduction}

\begin{figure*}[t]
\centering
\includegraphics[width=.85\textwidth]{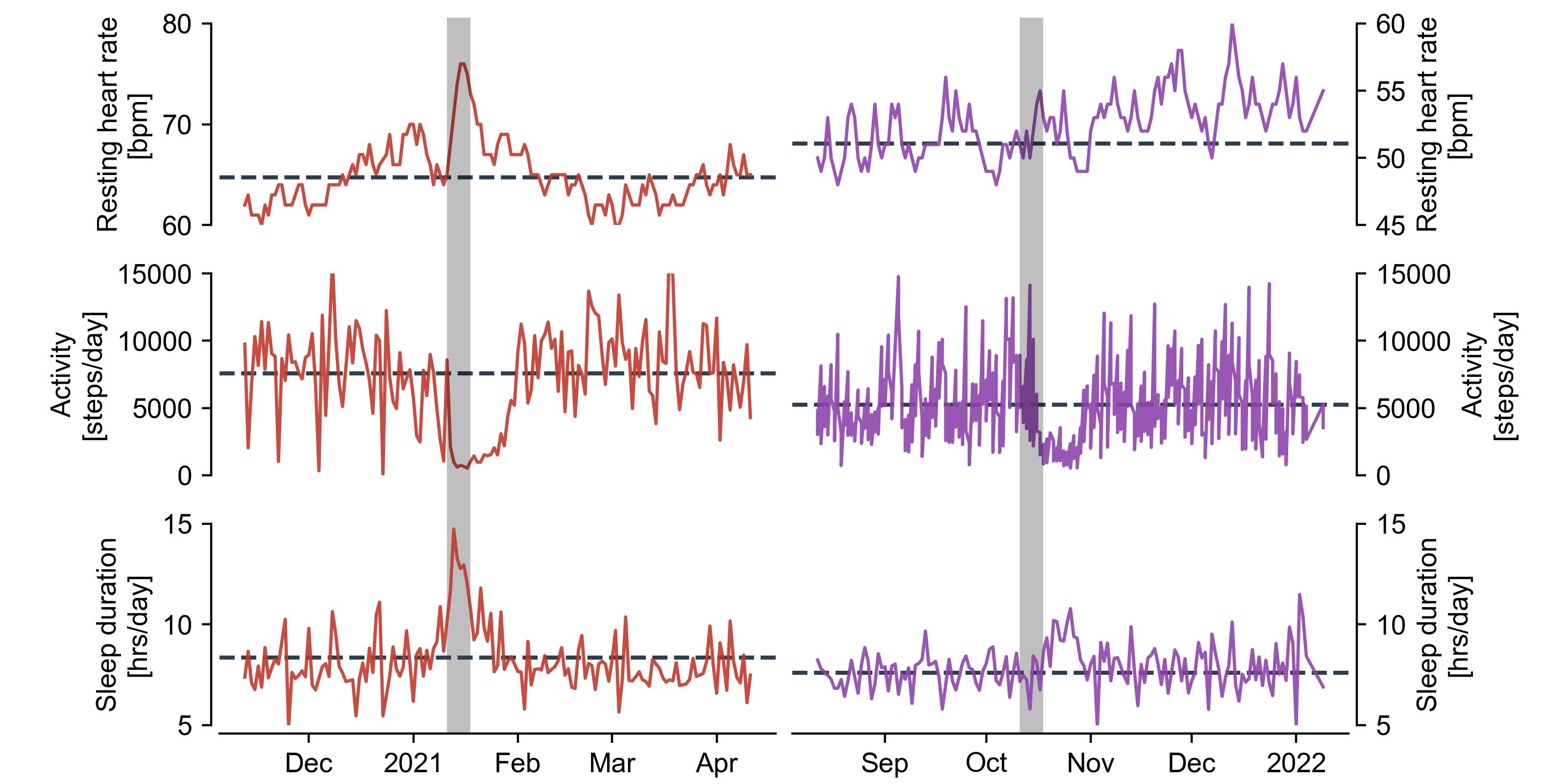}
\caption{Exemplary time series of daily resting heart rate (RHR, top), 
physical activity (middle) and sleep duration (bottom) for a time window of 150 days in a representative individual who was unvaccinated (left) and vaccinated (right) at the time of taking a positive PCR-test (grey shading).
Dashed lines denote the user's baseline, i.e., the average of the 60 days prior to the test. We observe a strong peak in RHR, a drop in physical activity and increased sleep duration for the unvaccinated individual around the time of the test. Similar patterns are observed for the vaccinated individual with respect to physical activity and sleep duration, the latter change being less pronounced. A visible change of RHR is absent for the vaccinated individual around the week of the test.}
\label{fig:example}
\end{figure*}

%General intro
COVID-19, the disease caused by infection with SARS-CoV-2, is usually accompanied by symptoms such as fever, cough, sore throat, shortness of breath, and fatigue~\cite{singhal_review_2020}. These symptoms are most prevalent during the acute phase of the disease, commonly defined as the four weeks following symptom onset~\cite{nalbandian_post-acute_2021}. 

%Post-acute COVID/Long-COVID
However, in some instances, these symptoms, along with a wide range of others, can persist, develop, or recur for weeks to months beyond this acute phase~\cite{wynberg_evolution_2021, michelen_characterising_2021} which is known as post-acute sequelae of SARS-CoV-2 infection (PASC) or, more commonly, Long COVID~\cite{wynberg_effect_2022, soriano_clinical_2021, nalbandian_post-acute_2021, greenhalgh_management_2020, datta_proposed_2020}. For instance, results from the UK-based REACT-2 study indicate that 15\% of surveyed people reported at least three COVID-19 related symptoms lasting for 12 weeks or more, while 38\% reported at least one~\cite{whitaker2021persistent, Nisreen2021road}. A follow-up study of COVID-19 patients discharged from a hospital in Wuhan, China revealed that 76\% of people reported at least one symptom 6 months after infection, 63\% of which specifically reported fatigue~\cite{huang_6-month_2021}, one of the most prevalent Long COVID symptoms~\cite{sudre_attributes_2021, carfi_persistent_2020, nalbandian_post-acute_2021}.

Other Long COVID symptoms include cognitive dysfunction, confusion, and brain fog as well as chest pain, shortness of breath, head- and muscle aches, dizziness, and heart palpitations~\cite{ziauddeen2022characteristics, sudre_attributes_2021}. They can be experienced by all age groups and are not exclusive to people with a severe course of the disease in its acute phase~\cite{Nisreen2021road}, though some indication exists that the risk of contracting Long COVID can be linked to the number and severity of symptoms experienced at the start of the illness~\cite{ziauddeen2022characteristics}. Generally, such long-term effects are not unique to COVID-19, but have also been reported for Middle East Respiratory Syndrome (MERS) and Severe Acute Respiratory Syndrome (SARS)~\cite{ong_pulmonary_2004, moldofsky_chronic_2011, hui_impact_2005, ahmed_long-term_2020}.

%Vaccines \& breakthrough infections
Vaccinations are effective against infection with SARS-CoV-2, hospitalization, ICU admission, and death following COVID-19 illness~\cite{tregoning_progress_2021, higdon_systematic_2022, maier_germanys_2021}. However, numerous reports of breakthrough infections after vaccination, especially for the two variants of concern B.1.617.2 (Delta) and B.1.1.529 (Omicron), have raised public concern~\cite{nixon2021vaccine}. Still, a large UK-based cohort study reported reduced risk of hospitalization or having more than five symptoms in the first week of illness after receiving at least one vaccination dose~\cite{antonelli_risk_2022}. Almost all symptoms were reported less frequently for breakthrough infections compared to infected unvaccinated individuals and vaccinated individuals were more likely to be fully asymptomatic~\cite{antonelli_risk_2022}. Vaccinations have also been shown to reduce the need for emergency care/hospitalization following breakthrough infection nearly ten-fold in a US population in Michigan~\cite{bahl_vaccination_2021}. Similar results were obtained from studies in Qatar~\cite{butt_outcomes_2021}, Spain~\cite{cabezas_associations_2021}, Italy~\cite{mateo-urdiales_initial_2021} and Israel~\cite{glatman-freedman_effectiveness_2021}, which consistently showed substantial reductions in severe cases between 38\% and 91\% percent. Likewise, high vaccine efficacies between 72\% against the B.1.351 variant (Beta)~\cite{abu-raddad_effectiveness_2021} and over 90\% for the B.1.617.2 variant (Delta)~\cite{glatman-freedman_effectiveness_2021,lopez_bernal_effectiveness_2021} were reported even though immunity waned across age groups a few months after the second vaccine dose~\cite{goldberg_waning_2021, haas_impact_2021}.

%Vaccines and long-covid
Although vaccines lower the risk of symptomatic and severe cases of breakthrough COVID-19 infection, it remains a subject of contention in COVID research whether they also attenuate
or prevent symptoms associated with PASC/Long COVID~\cite{ledford_vaccines_2021}. 
For instance, in an unrepresentative sample of 1,949 participants in a Long COVID Facebook group poll, 24 people reported Long COVID symptoms after symptomatic breakthrough infections~\cite{massey_breakthrough_2021}. A study of 1,500 people in Israel found that 19\% of recorded breakthrough cases resulted in symptoms that lasted longer than 6 months~\cite{bergwerk2021covid} and a study of around 6,000 adolescents found increased risk of prolonged symptoms even when people were fully vaccinated~\cite{stephenson_long_2022}. However, the comparatively small sample sizes in these studies effectively hinder general conclusions to be drawn~\cite{ledford_vaccines_2021}. Among studies with larger sample sizes, an analysis of 1,240,009 users of the UK-based {\em COVID Symptom Study app} in fact revealed a reduced risk of developing symptoms that last more than 28 days following a second vaccination dose~\cite{antonelli_risk_2022}. This finding is contradicted by another retrospective study of 10,024 vaccinated COVID-19 positive individuals in the UK that found no significant reduction in symptoms related to Long COVID~\cite{taquet_six-month_2021}. 

% Purpose of this paper / research question
Here, we provide evidence that vaccination against COVID-19 may significantly reduce the likelihood of developing long-term symptoms following an infection with SARS-CoV-2. We use large scale, daily data on resting heart rate, physical activity and sleep collected over a period of more than two years, see Fig.~\ref{fig:example}. This data was collected as part of the Corona Data Donation Project (Corona-Datenspende-App (CDA), \footnote{\protect\url{https://corona-datenspende.de}}), a smartphone app that allows users to submit such vital data by linking with consumer-grade wearable sensors, such as smartwatches and fitness trackers. The app was developed at the Robert Koch Institute, Germany's federal public health institute and released in Germany on April 12, 2020 to participants of age 16 and older.

As of April 3, 2022, a total of 1,177,636 people installed the CDA and 524,761 users submitted at least one data point. In March 2022, the app had approximately 190,000 monthly active users, and more than 120,000 people have submitted daily data over more than 600 days. Additionally, users can optionally participate in in-app surveys, which include questions regarding diagnostic test results and vaccination data. The large sample size, high temporal resolution and comparatively long observation period permits continuous tracking of an individual's biometrics and enables a fine-grained analysis of vital signs prior to a SARS-CoV-2 infection, throughout the acute phase of the disease, and during the recovery process. As such, the combination of survey and vital data enables analysis of long-term physiological and behavioral changes in COVID-19 positive and -negative individuals stratified by vaccination status. 

The systematic use of large-scale data collected via commercially available fitness trackers and smartwatches is a rapidly growing line of data-driven, medical research~\cite{colombo_current_2019, jaiswal_association_2020, quer_inter-_2020}. Wearable sensor data has been applied to study physiological markers of depression~\cite{ghandeharioun_objective_2017, wang_tracking_2018}, characterize daily physiology and circadian rhythms~\cite{bowman_method_2021}, and improve surveillance of influenza-like illness~\cite{radin_harnessing_2020}. In the COVID-19 context, this approach has been applied to the early detection of COVID-19 in individuals~\cite{mishra_early_2020}, predicting overall case numbers and changes in trends~\cite{zhu_learning_2020}, and discriminating COVID-19 positive from negative individuals~\cite{quer_wearable_2021, gadaleta_passive_2021}. Moreover, studies conducted prior to sufficient availability of vaccination data were able to link infections with SARS-CoV-2 to elevated resting heart rate that only returned to baseline levels an average of 79 days after symptom onset~\cite{radin_assessment_2021}.

\begin{table*}[t]
\centering
\begin{tabular}{l | r r r | r}
& Vaccinated & Unvaccinated & Negative & Total\\ [0.5ex] \hline\hline
Female & 901 (40.22\%) & 130 (40.88\%) & 2146 (38.74\%) & 3177 (39.24\%) \\ 
Male & 1330 (59.38\%) & 185 (58.18\%) & 3349 (60.46\%) & 4864 (60.07\%) \\ 
Other & 9 (0.40\%) & 3 (0.94\%) & 44 (0.79\%) & 56 (0.69\%) \\ 
Age (mean) & 47.09yr  & 47.97yr  & 50.09yr  & 49.17yr  \\ 
Age (std) & 11.37yr  & 11.73yr  & 12.37yr  & 12.15yr  \\ 
\end{tabular}
\caption{Number of users per gender as well as mean and standard deviation of age in the respective cohorts under study. Percentages in parentheses indicate the respective share of users within the specific cohort.}
\label{tab:user_stats}
\end{table*}

\section{Results}

We computed weekly changes in resting heart rate (RHR), daily steps, and sleep duration around the date of a COVID-19 PCR-test for a total of 8,097
individuals, of which 2,240 experienced a breakthrough infection, 318 were infected prior to achieving immunity, and 5,539 reported negative PCR-tests, thus serving as a control group (see SI, Fig.~S6). The per-user baseline against which changes are computed is given by the respective average of each variable over the 8 weeks preceding a confirmed infection with SARS-CoV-2, indicated by the approximate date of a PCR-test (given at a weekly temporal resolution). Signals were normalized by subtracting the daily average from all individual time series to account for seasonal effects, e.g. naturally prolonged sleep duration in winter and increased activity in summer. Thus, individual vital data is always measured relative to the population-wide average on a given day. Further details on the characteristics, pre-processing, and analysis of the data are provided in the Materials \& Methods Section. 

\subsection{Vaccinations mitigate long-term vital changes in COVID-19 positive individuals}
\label{sec:results_long_term}

\begin{figure}[t]
\centering
\includegraphics[width=.9\linewidth]{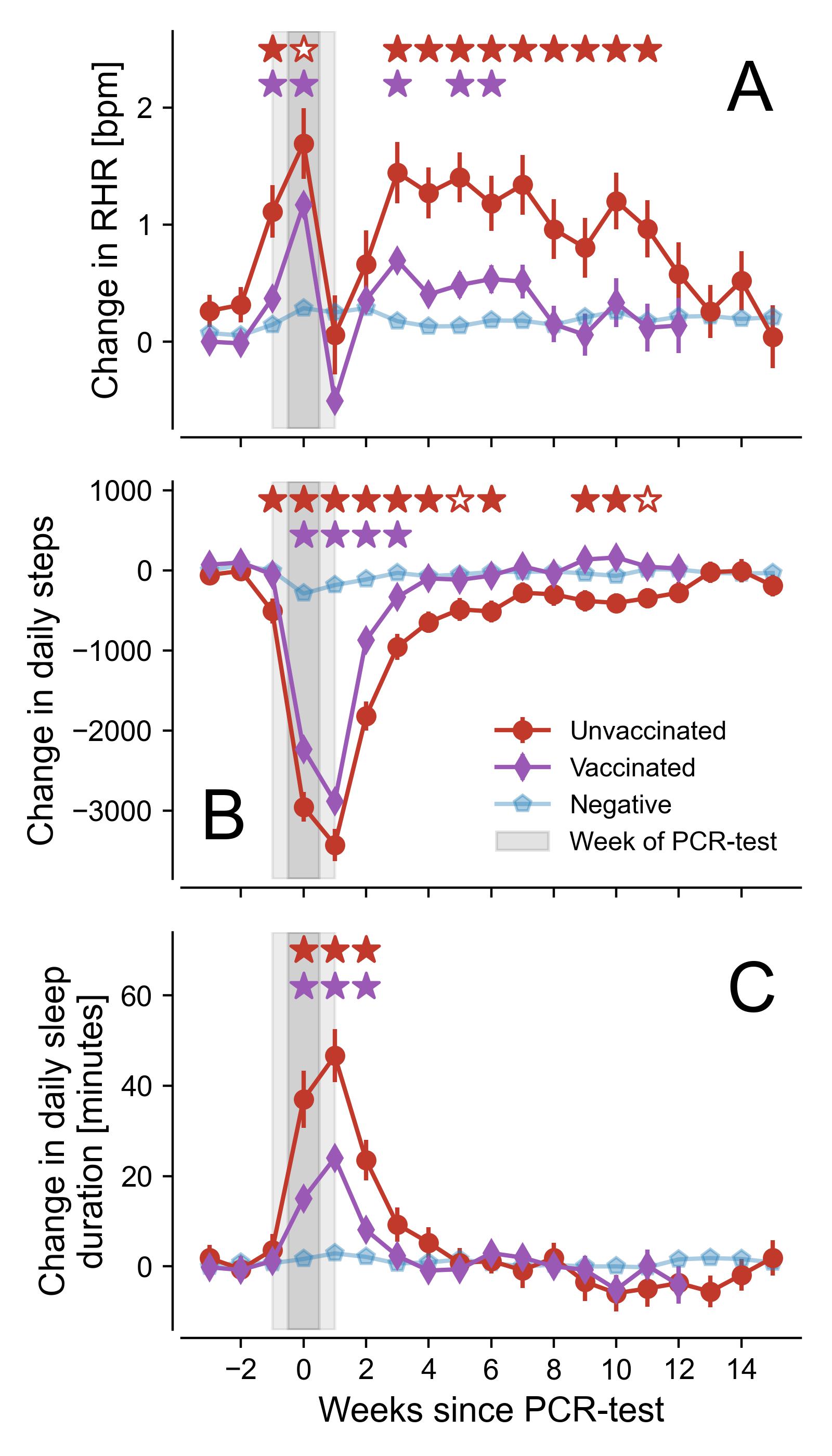}
\caption{Changes in RHR, activity, and sleep duration in unvaccinated infected (red) and vaccinated infected (purple) as well as negative controls (blue). Changes are measured relative to the two months preceding the test. Errors bars indicate standard error. Filled (empty) red stars indicate periods where the average vital change of unvaccinated individuals is stronger than that of vaccinated (negative) individuals using a one-sided Welch t-test with a significance level of $\alpha=0.01$. Purple stars indicate significant differences between vaccinated and COVID-19 negative individuals. }
\label{fig:long_covid}
\end{figure}

\begin{figure*}[t]
\centering
\includegraphics[width=.95\textwidth]{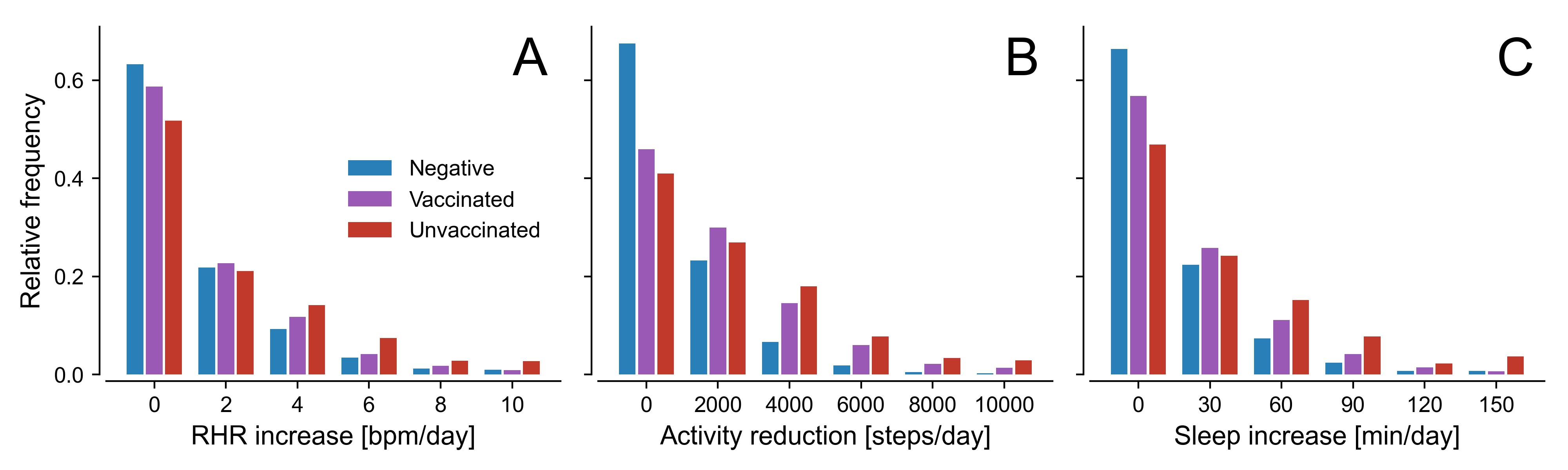}
\caption{Relative frequency of weekly average vital changes in weeks zero to four following a PCR-test for vaccinated and unvaccinated COVID-19 positive individuals as well as COVID-19 negative individuals. Numbers on the vertical axis indicate the center of each bin. Values outside the specific bins are added to the smallest and largest bin, respectively.}
\label{fig:distributions}
\end{figure*}

%Corona-Datenspende, loosely translated as Corona Data Donation, of the Robert Koch-Institute is an app-based, longitudinal research study that enrolls participants of in Germany. 
We first evaluated the evolution of average changes in weekly RHR, step count, and sleep duration in the weeks following a PCR-test separately for each user cohort, depicted in Fig~\ref{fig:long_covid}. 

On average, the RHR of unvaccinated users with SARS-CoV-2 infection increased by ${\sim} 1.7$ beats per minute in the week of the PCR-test and only returned to baseline levels after 11 weeks (Fig.~\ref{fig:long_covid}A). This finding qualitatively confirms similar results obtained in earlier studies~\cite{radin_assessment_2021, natarajan_occurrence_2022} that did not specifically differentiate by vaccination status. We found a pronounced drop in RHR at around one week after a PCR-test with values that decreased even below baseline for vaccinated users. This aligns with earlier studies~\cite{radin_assessment_2021} and potentially indicates transient bradycardia following infection~\cite{amaratunga_bradycardia_2020}. RHR of unvaccinated individuals already increased significantly in the week preceding a PCR-test at values of ${\sim} 1.1$ beats per minute above normal. We found weaker average deviations in RHR for vaccinated individuals with a maximum value of ${\sim} 1.2$ beats per minute in the week of the PCR-test. These deviations were accompanied by a swifter return to baseline levels after approximately 3 to 6 weeks. Except for the two weeks following a PCR-test the average RHR-change for vaccinated individuals was approximately two to three times lower than for those that were unvaccinated (Fig.~\ref{fig:long_covid}A), potentially indicating a milder course of the disease on average. 

The average daily activity (Fig.~\ref{fig:long_covid}B) decreased in the week of the PCR-test by ${\sim} 2,000$ and ${\sim} 3,000$ steps per day for vaccinated and unvaccinated individuals, respectively. For both groups, the reduction usually began the week of a positive PCR-test, and thus might have been partially modulated by changes in behavior, i.e. self-isolation. A return to baseline activity among vaccinated individuals occurred after only 4 weeks compared to around 6 to 11 weeks for a small subset of unvaccinated users, indicating that at least some suffered from a prolonged reduction in activity, thereby skewing the mean towards negative values.

Finally, the average sleep duration of unvaccinated individuals increased abruptly by ${\sim} 37$ minutes per day during the week of the PCR-test. For vaccinated individuals, this effect was reduced by more than half to an average of only ${\sim} 15$ minutes per day. Similar magnitudes were observed in the first week following a PCR-test, where sleep duration was increased by ${\sim} 46$ and ${\sim} 24$ minutes per day for unvaccinated and vaccinated users, respectively. Sleep duration returned to baseline values quickly for both user groups as compared to activity or RHR. By the third week, anomalies in sleep duration were comparable with that of the COVID-19 negative control group. Still, we found a significant increase of approximately two times in the average need for rest during the acute phase of the disease when comparing vaccinated and unvaccinated individuals. 

We also found small variations in RHR, activity, and sleep duration in COVID-19 negative individuals around the test period, which might be caused by other diseases, such influenza or a common cold, that might have caused people to take a PCR-test in the first place.

\subsection{Distribution of extreme vital changes in the acute phase}
As the most prominent changes were observed during the acute phase of the disease, cf.\ Fig.~\ref{fig:long_covid}, we next investigated the distributions of weekly increases in RHR and sleep duration and decreases in activity for each cohort in the four weeks following a PCR-test, Fig.~\ref{fig:distributions}.

\begin{figure}[t]
\centering
\includegraphics[width=.9\linewidth]{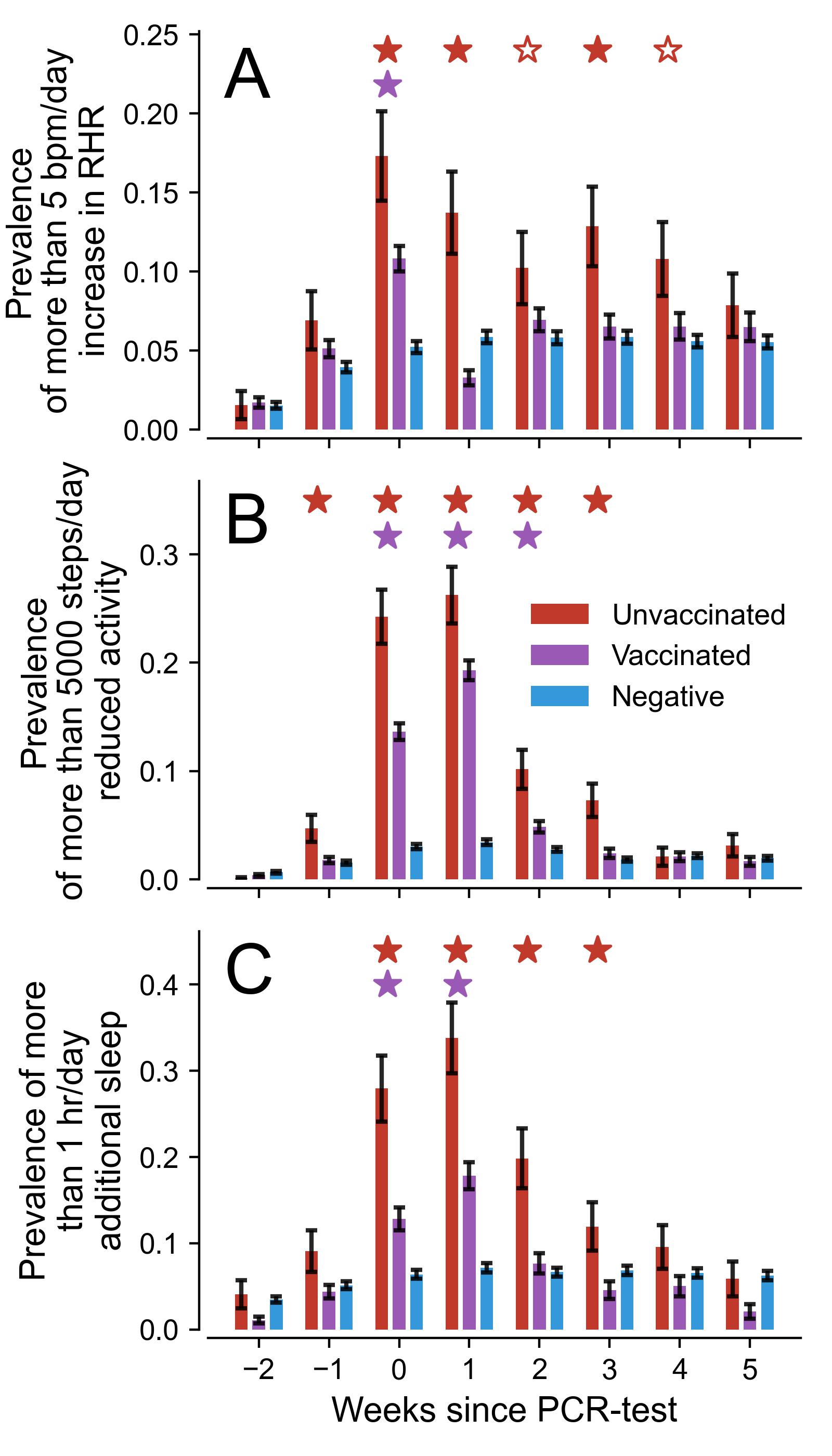}
\caption{Share of donors in each cohort whose weekly average vital data exceeded a specified threshold. (A) Share of donors with more than 5 bpm/day RHR increase. (B) Share of donors with an activity reduction of more than 5,000 steps/day. (C) Share of donors with an increased sleep duration of more than 1 hour/day. Red, purple and blue bars indicate unvaccinated, vaccinated and COVID-19 negative donors, respectively. Error bars indicate the standard error of a binomial distribution. Filled (empty) red stars indicate periods where the respective prevalence in unvaccinated individuals was stronger than that in vaccinated (negative) individuals using a one-sided two proportion z-test with a significance level of $\alpha=0.01$. Purple stars indicate significant differences between vaccinated and COVID-19 negative individuals. }
\label{fig:acute_phase}
\end{figure}

For all metrics, the frequencies of changes decayed continuously with increasing values. As expected, the smallest changes (less than 1 bpm/day of RHR change, less than 1,000 steps reduction in activity/day, and less than 15 minutes of additional sleep/day) were most commonly observed in the COVID-19 negative cohort. Likewise, between 40\% and 50\% of all COVID-19 positive individuals experienced only small changes in all three metrics, likely indicating comparatively mild courses. Note that these numbers are well below the commonly reported percentage of mild and moderate cases of at least 80\%~\cite{wu_characteristics_2020, schilling2021verschiedenen} while exceeding rough estimates (${\sim}41\%$) for asymptomatic infections in confirmed COVID-19 cases. Hence, we found a reasonable amount of individuals with little or no changes in their vital data during SARS-CoV-2 infection. The observed frequencies of larger and more extreme vital changes in the unvaccinated cohort were consistently higher than those measured for vaccinated and COVID-19 negative individuals, again indicating increased likelihood for a severe course in unvaccinated individuals, Fig.~\ref{fig:distributions}.

\subsection{Vaccinations reduce short-term risk of severe vital changes in COVID-19 positive individuals}
\label{sec:results_extreme}

In order to clarify the prevalence of severe courses in acute cases, we measured how such extreme changes in vital signals were distributed for the unvaccinated, vaccinated and COVID-19 negative cohorts in the first weeks after taking a PCR-test. The results are depicted in Fig.~\ref{fig:acute_phase}. Specifically, we computed within each cohort the share of users whose vital changes exceeded a certain threshold. Specifically, we counted individuals whose change in RHR exceeded 5 bpm/day, which is indicative of an increase of half a degree in body temperature~\cite{karjalainen_fever_1986}. Moreover, we chose thresholds to define extreme activity reduction of 5,000 steps/day as well as pronounced sleep prolongation of more than 1 hour per day, both of which well exceed the maximum observed average change in Fig.~\ref{fig:long_covid} while still yielding reasonable frequencies in Fig.~\ref{fig:distributions}. 

In the unvaccinated group, the frequency of more than 5 bpm/day RHR change varied between 17.5\% in the week of and 10\% in the fourth week after a positive PCR-test, Fig.~\ref{fig:acute_phase}A. For weeks zero, one, and three those frequencies were significantly larger than those of the vaccinated cohort. The same held true for weeks two and four when comparing the unvaccinated cohort with the COVID-19 negative control group. Only from week five onward were extreme RHR changes equally likely in COVID-19 positive individuals as in the negative control cohort. In contrast, in the vaccinated cohort, extreme RHR changes were only significantly more common than controls during the week of the PCR-test. However, we note that, even in the week of the PCR-test, the respective frequency was still only approximately half as large as for the unvaccinated cohort. 

Likewise, we found significantly increased prevalence of drastic reduction in activity for both the vaccinated and unvaccinated cohorts in the first 2-3 weeks following a PCR-test when compared to the COVID-19 negative control group, Fig.~\ref{fig:acute_phase}B. In addition, the prevalence in the unvaccinated cohort was significantly larger compared to the vaccinated group in all significant weeks, again indicating a reduced risk of severe illness following full vaccination. Moreover, a small share (${\sim}5\%$) of unvaccinated individuals already showed substantial activity reduction in the week prior to taking a PCR-test, indicating a potential precursor for the developing disease. No significant extreme reductions in activity were observed after the third week, Fig.~\ref{fig:acute_phase}B. 

Finally, we considered the frequency of individuals with an increased sleep duration of 1 hour/day, indicative of a strongly increased need for rest in the acute phase of COVID-19, Fig.~\ref{fig:acute_phase}C. Both COVID-19 positive cohorts showed greatly increased frequency in sleep prolongation during weeks zero and one following a PCR-test, with more than 30\% of cases in the unvaccinated group and 10-20\% of cases in the vaccinated cohort. Extended sleep duration was also significantly prominent in the unvaccinated cohort in the second week after a positive PCR-test with a prevalence of more than 20\% compared to less than 10\% in the vaccinated and the control cohorts. Hence, roughly 3 to 4 out of 10 unvaccinated individuals experienced an increased sleep duration of more than 1 hour/day for an extended period of 2 to 3 weeks. 

\section{Discussion \& Conclusion}
\label{sec:discussion}
We analyzed changes in resting heart rate (RHR), physical activity, and sleep duration around the time of a PCR-test for 2,240 vaccinated and 318 unvaccinated COVID-19 positive individuals as well as 5,539 individuals in a COVID-19 negative control group. Participants in this study were self-recruited, often following media announcements. They submitted their vital data and meta-information, i.e., socio-demographics, PCR-test dates and results, and vaccination status, via the Robert Koch Institute's Corona-Datenspende smartphone app (CDA), downloadable free of charge for German residents over the age of 16. 

We found that average deviations and subsequent stabilizations in vital signals were most pronounced for unvaccinated individuals, with the longest normalization period spanning an average of 11 weeks post PCR-test week for both RHR and activity. Similar findings have been obtained in other studies that, although not explicitly stated, likely mostly considered unvaccinated individuals due to the scarcity of vaccines at the time~\cite{radin_assessment_2021, natarajan_occurrence_2022}. Average vital changes for vaccinated persons were less pronounced, albeit at times still significantly different from the COVID-19 negative control group. In addition to prolonged average changes, we found that extreme values were more likely to be observed for unvaccinated individuals in the acute phase of the disease when compared to vaccinated individuals or the negative cohort. Finally, we observed that both RHR as well as the step count of unvaccinated COVID-19 positive individuals, already differed significantly from the negative control group in the week prior to taking a PCR-test, hinting at its potential to serve as an early warning indicator of a coming illness~\cite{radin_harnessing_2020, mishra_early_2020}. 
Our results provide further evidence that vaccinations can not only mitigate severe cases of acute COVID-19, which is in line with the broader literature~\cite{tregoning_progress_2021, higdon_systematic_2022, maier_germanys_2021, antonelli_risk_2022, bahl_vaccination_2021, butt_outcomes_2021, abu-raddad_effectiveness_2021, cabezas_associations_2021, mateo-urdiales_initial_2021}, but also highlight their potential for attenuating long-term physiological and behavioral changes~\cite{ledford_vaccines_2021,massey_breakthrough_2021,stephenson_long_2022, taquet_six-month_2021}. Our results exemplify the great potential that lies in passive sensing for public health research as it fosters robust and large-scale analysis of long-term, high-resolution longitudinal data with interpretable metrics that can be linked to an individual's physiology~\cite{cornet_systematic_2018, trifan_passive_2019}.  

Our analysis comes with some limitations that need to be considered when contextualizing the above results. First and foremost, our analysis did not discriminate infections by the respective variant of concern (VOC) that was predominant at the time a PCR-test was taken. Furthermore, individuals that were designated as unvaccinated were primarily assigned to this cohort because, at the time of infection, vaccine availability was limited (see SI Sec.~I). Hence, all unvaccinated donors were likely infected with B.1.1.7 (Alpha) or the wild-type which might have triggered different physiological responses than the later emergent variants, specifically B.1.617.2 (Delta) and, more recently, B.1.1.529 (Omicron). The majority of breakthrough infections in our data set were reported when these latter two variants were prevalent, meaning that the observed effects could partially also be explained by weaker physiological responses to Omicron~\cite{chen_omicron_2022}. To account for this effect, we performed an additional sensitivity analysis (see SI Sec.~IIA) that only considered infections reported prior to December 15, 2021 and thus likely caused by Delta, which has been reported to cause more severe cases than Alpha~\cite{ong_clinical_2021, fisman_evaluation_2021, sheikh_sars-cov-2_2021}. When only considering this subset of users in the vaccinated cohort, we found that the results presented in Fig.~\ref{fig:long_covid} still held, indicating that vaccinated individuals likely infected with the Delta variant exhibited significantly weaker average changes in vital data compared to the unvaccinated group. We also compared average vital changes in that same cohort of vaccinated individuals infected before December 15, 2021 to individuals who reported infections after that date and, therefore, were likely infected with the Omicron VOC. We found hardly any significant differences in the temporal evolution of vital changes between the cohorts (see SI Sec.~IIB). Hence, in the context of our analysis, it is reasonable to combine all recorded breakthrough infections into a single cohort for ease of interpretability and without having results skewed by the influence of a single variant of concern.

By now, almost all users of the Corona Data Donation Project are at least fully vaccinated or have received a booster vaccination. Physiological responses to more recent variants of concern in unvaccinated individuals could, therefore, only be recorded if unvaccinated individuals were specifically recruited. In addition, we did not explicitly account for the time between the receipt of the latest vaccination dose and the date of breakthrough infection, which ignores the potential effects of waning immunity~\cite{de_gier_covid-19_2021, tartof_effectiveness_2021}. However, most breakthrough infections in our data set were recorded in the first four months after achieving immunity (see SI Sec.~III), likely indicating sufficient protection in the majority of the vaccinated cohort. 

We also note that neither of our three cohorts is representative of the German population. Our sample shows a large over-representation of male individuals (see Table~\ref{tab:user_stats}), as well as an under-representation of adolescent and elderly (65+) persons, see SI Sec.~IV for the distribution of age groups. Moreover, there is good reason to assume that our study population is more health-conscious than the average population since the usage of fitness trackers is partially correlated with or, at least, facilitates awareness of health-related behavior \cite{wu_how_2016}. Likewise, the cohorts might not be fully representative of one another, even though the basic proportions of gender and age match well across them, Table~\ref{tab:user_stats}. There are many other additional confounding factors that might influence the observed vital changes, such as pre-existing conditions, self-reported symptoms during the disease, socio-economic status, as well as demographics including age, sex, and body mass index. Due to the limited sample size, we refrained from performing any further discrimination along these potentially confounding factors, but such analyses could be performed in the future if the recorded cohorts increased in size. 
However, despite the above limitations, our analysis provides relevant insights regarding the efficacy of vaccine against long-term effects of COVID-19. Because vulnerable groups, such as the elderly, are under-represented, the observed differences might become more pronounced if more people from such groups participated in the study. As such, the results for unvaccinated individuals might indicate a lower bound for expected vital changes which might become larger when a more representative cohort is considered, thereby potentially increasing the observed differences between our cohorts further. 

Furthermore, we acknowledge that we cannot disentangle whether changes in vital data, especially activity and sleep duration, were caused by altered behavioral in response to a positive diagnosis or whether those changes were an actual physiological imprint of an acute infection. While we likely observed a combination of both effects, it remains impossible to quantify their individual influence. However, we may assume that the mere effect of isolation reduces the opportunity for physical activity equally for unvaccinated and vaccinated individuals, thereby making it an unlikely explanation for all the observed changes in daily activity. Still, across all metrics, the average deviation in step counts were most similar between the COVID-19 positive cohorts, indicating that these changes are, in fact, partially driven by self-isolation.    

Future work should aim to reproduce and validate the results obtained in this study, ideally with data that is collected in a similar fashion such as through the {\em DETECT}~\cite{radin_hopes_2021} or {\em Evidation}~\cite{shapiro_characterizing_2021} systems in the US as well as {\em TemPredict} which covers a broad range of international users~\cite{mason_detection_2022, smarr_feasibility_2020}. We further propose to investigate and improve the representativeness of the user sample, i.e., by comparison with common health survey programs, such as GEDA in Germany~\cite{damerow_developments_2020} or NHIS in the US~\cite{national1986national}. One should then aim to specifically advertise for an increased participation of currently under-represented groups, potentially by providing wearable devices to users that could normally not afford such devices and are therefore missing from the data set. We suggest to also incorporate higher-frequency data recorded with a temporal resolution on the scale of minutes~\cite{bowman_method_2021}. Such data would allow for the quantification of more subtle changes in physiology, such as the postural orthostatic tachycardia syndrome (POTS)~\cite{low_chapter_2012}, another typical condition associated with Long COVID~\cite{raj_long-covid_2021, miglis_case_2020}. After all, it is a unique advantage of wearable sensors that data can be measured over extended periods at high resolution and minimal burden to the individual, thereby making them a promising tool to complement traditional clinical methods for a data-driven approach to public health research~\cite{radin_hopes_2021, trifan_passive_2019, cornet_systematic_2018}. 

\begin{acknowledgments}
We thank Michael Hallek, Christian Karagiannidis, and Christa Scheidt-Nave for helpful discussions during the preparation of this manuscript. Paul Burggraf and Hannes Schenk are greatfully acknowledged for their technical support in the data collection process. We also thank Claudia Enge and Lorenz Wascher for their continuous assistance regarding data privacy and data protection.
\end{acknowledgments}

\section*{Material \& Methods}

\subsection*{Data Characteristics}
Between April 12, 2020 and April 3, 2022, a total of 524,761 people installed the Corona-Datenspende App and submitted at least one vital data point. Of these users, 38,853 people agreed to also participate in regular surveys about COVID-19 test results, vaccination status, and other information relevant for pandemic research. 29,323 users submitted their vaccination status and the months of receiving doses and 20,461 provided the week and result of their first positive PCR-test or their first ever PCR-test if all results were negative. The overlap between users who submitted both vaccination status and test results resulted in 16,693 people. 

\subsection*{Data Preprocessing}
Due to inconsistent measurements in sleep duration for Apple devices following a manufacturer update on October 10, 2022, 325 affected users were removed from the data set. In addition, users who achieved immunity from a single dose with the vaccine Ad26.COV2.S (Janssen) were excluded from the analysis (338 users). Of this subset, we kept users who donated at least one vital data point between eight weeks preceding and twenty weeks following their PCR-test (12,198 users). We computed weekly averages if at least six data points were present in a given week. Users with less than three weeks of sufficient data preceding their test were dropped since no reliable baseline could be computed, leaving a total of 8,097 users. We considered all users who received at least two vaccination doses prior to their positive PCR-test as {\em vaccinated} and all others {\em unvaccinated}, such that 2,240 individuals experienced a breakthrough infection, 318 were infected prior to achieving immunity, and 5,539 only reported negative PCR-tests. A detailed cohort-diagram can be found in the SI, Fig.~S6.

\subsection*{Data Analysis}
As a first step, we computed per-user anomalies of all vital data. For this purpose, we subtracted daily population-wide averages of resting heart rate, step count and sleep duration from each user's time series in order to account for seasonal effects, such as increased activity in summer or prolonged sleep duration in winter. For an increased accuracy, we used the data of all 16,368 users that are left in the study cohort after removing apple users with inconsistent data for this purpose (see above). After this transformation, a value of zero indicated that the user's vital measurement was en par with the population-wide average on a given day, while positive and negative values indicated above- and below-average values, respectively. 

Next, user data was down-sampled into weekly bins, enabling the same temporal resolution as the approximate date of reported PCR-tests. We then computed for each user and vital metric an individual baseline as the average over the 8 weeks prior to the PCR-test. We subtracted this baseline from each time series to obtain the corresponding deviations in vital signals. 

For all users, the time series were then aligned with the week of a PCR-test and averaged (for the results in Sec.~\ref{sec:results_long_term}) or thresholded (for the results in Sec.~\ref{sec:results_extreme}) depending on the desired analysis.

\subsection*{Statistical analysis}
For all discussions of differences in average vitals, we used a one-sided Welch t-test. For analyzing the prevalence of extreme vital changes, we applied a one-sided two proportion z-test. For both tests, we used a significance level of $\alpha=0.01$. One-sided tests were used since we put our focus on whether vital changes of unvaccinated individuals {\em exceeded} those of the vaccinated or negative cohorts. Likewise, we were only interested in vital changes of vaccinated individuals if they {\em exceeded} those measures observed for the negative control cohort.  

\subsection*{Ethical considerations}
All individuals participating in the Corona Data Donation Project provided informed consent electronically via the app. The study was reviewed and approved by the Data Privacy Officer at the Robert Koch-Institute (2021-009) in agreement with the Federal Commissioner for Data Protection and Freedom of Information (BfDI).

%\bibliography{library}

%\pagebreak
\newpage

\newpage

\onecolumngrid
\begin{center}
  \textbf{\large Evidence for positive long- and short-term effects of vaccinations against COVID-19 in wearable sensor metrics --- Supplementary information}\\[.2cm]
\end{center}

\setcounter{equation}{0}
\setcounter{section}{0}
\setcounter{figure}{0}
\setcounter{table}{0}
\renewcommand{\theequation}{S\arabic{equation}}
\renewcommand{\thefigure}{S\arabic{figure}}
\renewcommand{\thetable}{S\arabic{table}}

\section{Distribution of positive PCR-tests over time}

\begin{figure}[h!]
\centering
\includegraphics[width=.6\linewidth]{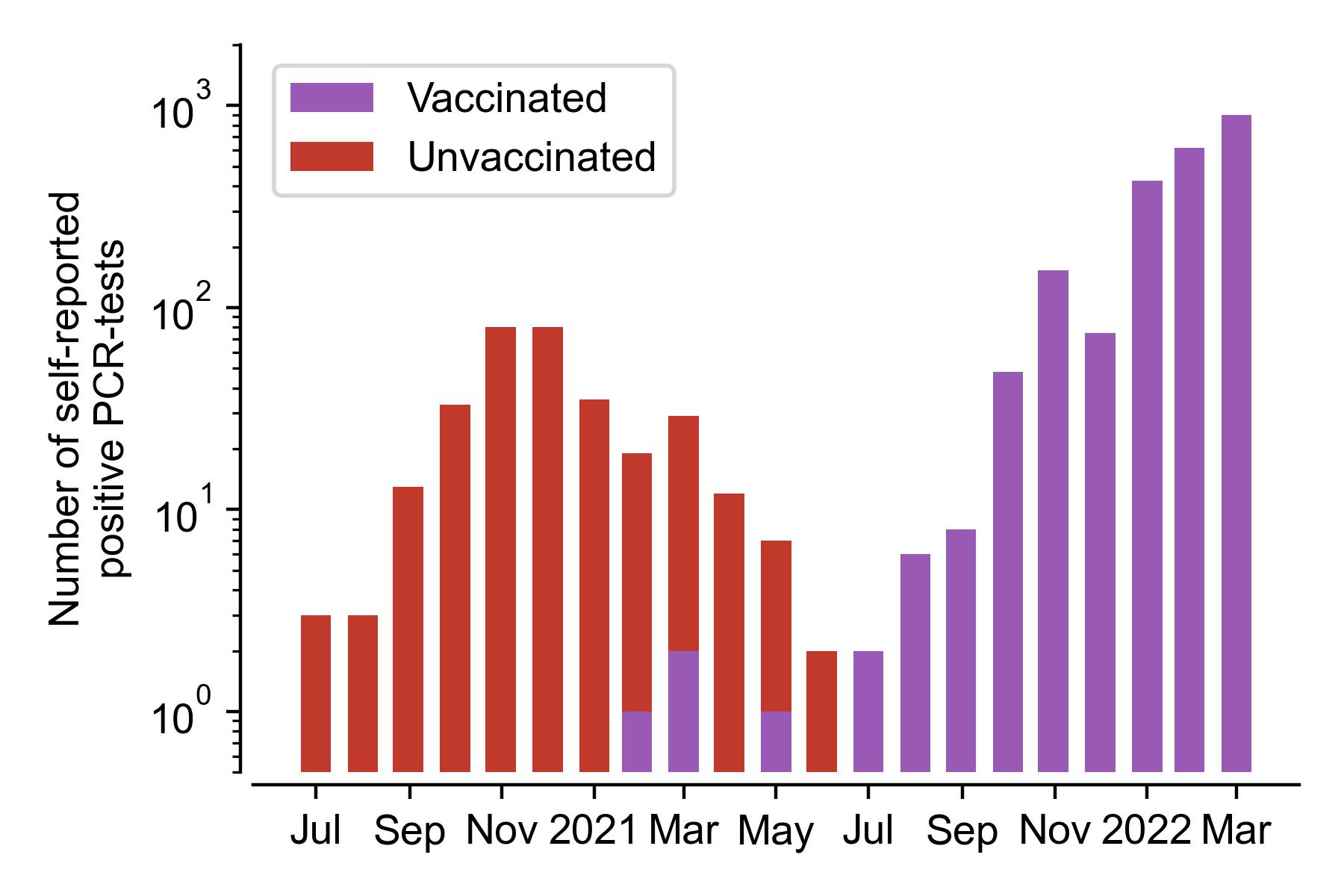}
\caption{Absolute count of self-reported positive PCR-tests per month for vaccinated individuals (purple) and unvaccinated individuals (red). Note the logarithmic scale due to an increasing number of positive tests from 2022 onwards.}
\label{fig:pcr_counts}
\end{figure}

As mentioned in the main manuscript most infections in vaccinated individuals took place during the first three waves of the pandemic while breakthrough infections are mostly recorded during the Delta and Omicron waves in late 2021 and 2022, Fig.~\ref{fig:pcr_counts}. In fact, the latest infection of an unvaccinated person was reported in June 2021 and only 4 vaccinated individuals report an infection prior to that month.

\section{Influence of variant-specific breakthrough infections on the main results}

\subsection{Vital changes after breakthrough infections with B.1.617.2 compared to unvaccinated individuals}
\label{sec:omicron1}

\begin{figure}[t!]
\centering
\includegraphics[width=.5\linewidth]{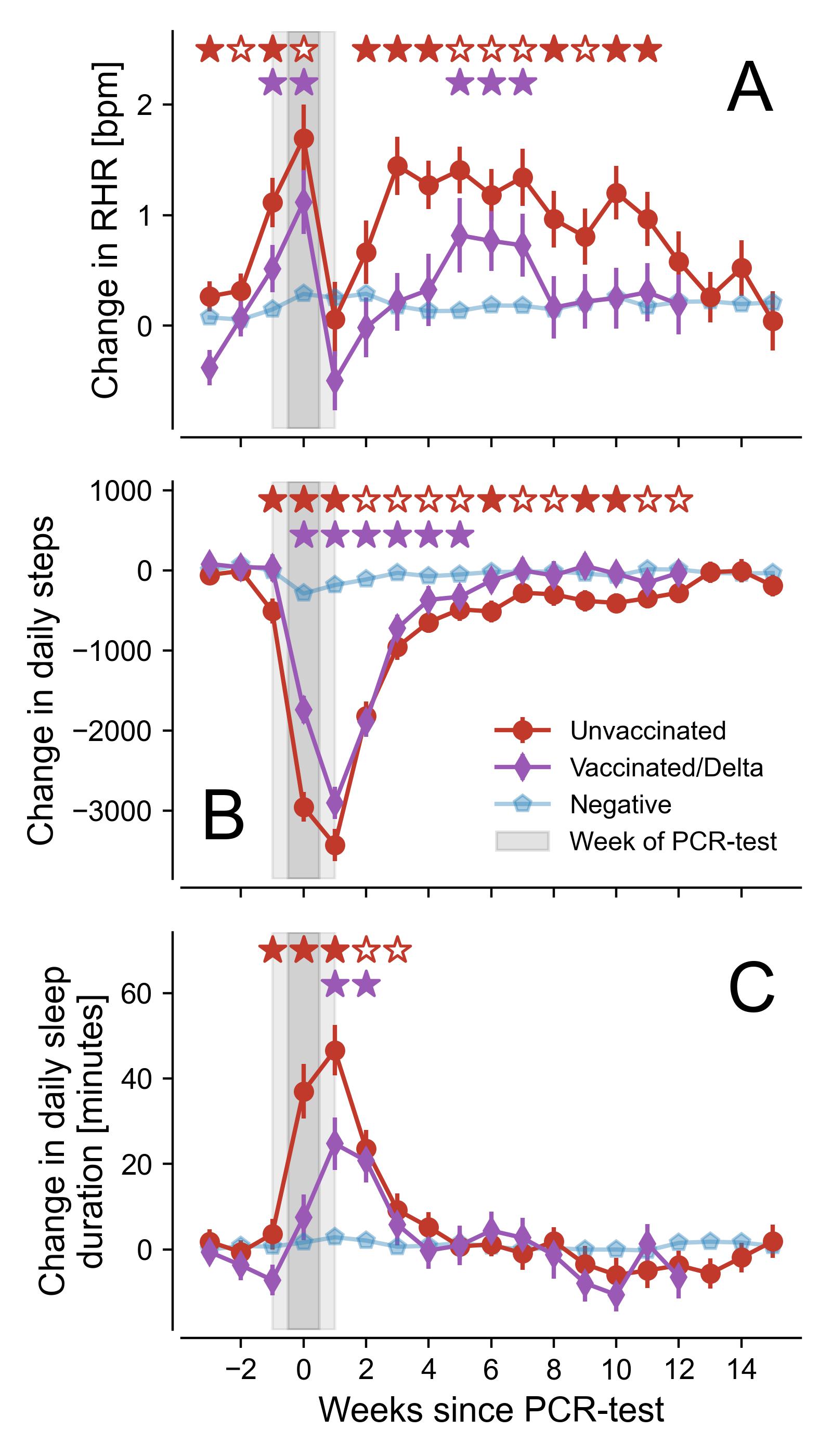}
\caption{Same as Fig.\ 2 in the main manuscript, but only considering breakthrough infections that took place before December 15, 2021, i.e., most likely caused by the B.1.1.7 (Delta) Variant. Filled (empty) red asterisks indicate periods where the average vital change of unvaccinated individuals is stronger than that of vaccinated (negative) individuals using a one-sided Welch t-test and a significance level of $\alpha=0.05$. Purple asterisks indicate significant differences between vaccinated and COVID-19 negative individuals.}
\label{fig:long_covid_delta}
\end{figure}

To account for the fact that the analysis in the main manuscript does not discriminate breakthrough infections by the respective variant of concern, we repeat the analysis in Sec.~2C only for cases that were reported before December 15, 2021. During that time only the more severe B.1.617.2 (Delta) Variant was predominant, Fig.~\ref{fig:long_covid_delta}~\cite{koch-institut_sars-cov-2_2022}.

Recall from the main manuscript, that we found significant differences in resting heart rate (RHR) between unvaccinated and vaccinated individuals at a high significance level ($\alpha=0.01$) at almost all weeks, except the two weeks following a positive PCR-test. Likewise vaccinated individuals differed significantly from the COVID-19 negative control group at weeks -1, 0, 3 and 5-6. In addition, we found that RHR-changes of vaccinated individuals consistently fall below those of unvaccinateds. If we now restrict our analysis to infections in vaccinated individuals that were likely caused by Delta, we find that this general pattern still holds, Fig.~\ref{fig:long_covid_delta}A. Due to the smaller sample size of Delta infections we do, however, adjust the significance level to $\alpha=0.05$. Average RHR-changes for unvaccinated individuals still significantly exceed those of vaccinateds in the week preceeding a positive PCR-test, as well as the weeks 2-4, 8 and 10-11 after the test. Hence, the general trends towards lower expected RHR-changes for vaccinated individuals also holds if the analysis is restricted to the Delta variant of concern. This is particular important in the context of our work since Delta is generally considered the variant that causes the most severe cases. Hence, it is reasonable to conclude that an infection of a vaccinated person with Delta is likely still less pronounced with respect to RHR changes than an infection with the B.1.1.7 (Alpha) variant or the wild-type of SARS-CoV-2, albeit the latter two being associated with less severe courses of the disease. 

We find similar patterns for physical activity, i.e. the weekly averaged number of daily steps taken, even though vaccinated individuals show significantly reduced values for weeks 0 to 5 (again using a significance level of $\alpha=0.05$) compared to only the first three weeks after a PCR-test for the entire cohort, cf.~Fig.~2B in the main manuscript. Still, activity reduction in unvaccinated individuals takes much longer (up to 12 weeks) to return to normal values, again indicating that vaccines also mitigate risks of long-term activity reduction after an infection with Delta. 

Ultimately, sleep duration (Fig.~\ref{fig:long_covid_delta}C) of vaccinated individuals after an infection with Delta is only significantly increased for the two weeks following a positive PCR-test which is en par with the observations for the entire cohort, compare again~Fig.~2C in the main manuscript. Hence, also with respect to this vital type we see no significant qualitative differences in the vaccinated cohort if we restrict our analysis to Delta infections.

\subsection{Vital changes for breakthrough infections with B.1.617.2 compared to B.1.1.529}
\label{sec:omicron2}

\begin{figure}[t!]
\centering
\includegraphics[width=.5\linewidth]{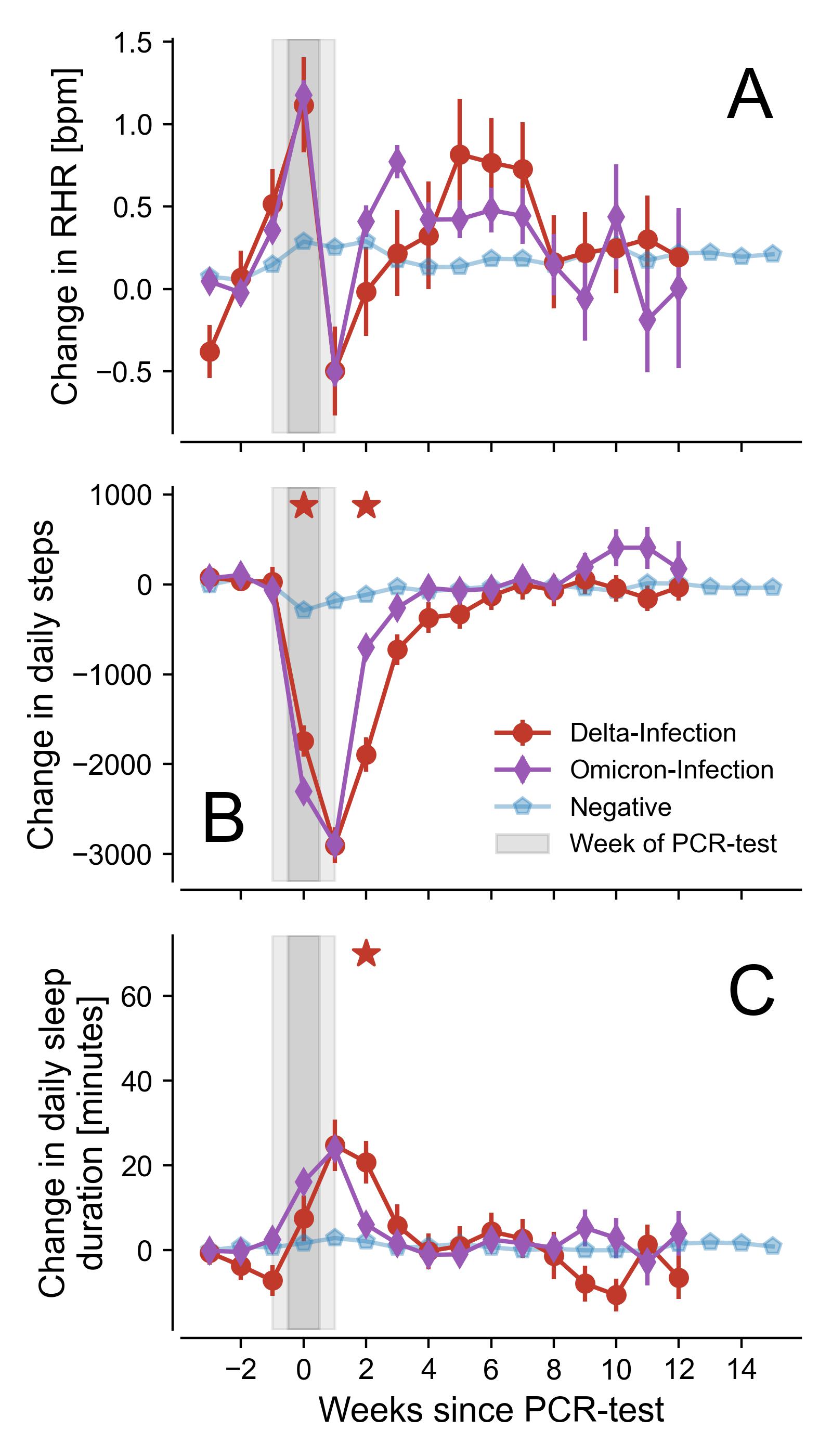}
\caption{Changes in RHR, activity, and sleep duration in two vaccinated user cohorts with recorded infections before December 15, 2021 (red) and after (purple) as well as negative controls (blue). Changes are measured relative to the two months preceding the test. Errors bars indicate standard error. Red stars indicate periods where the average vital changes in the two vaccinated cohorts are statistically different using a two-sided Welch t-test and a significance level of $\alpha=0.01$.}
\label{fig:delta_omicron}
\end{figure}

We perform an additional analysis to compare the observed vital changes in vaccinated individuals between the two pandemic waves for which respective test-dates are available, see also Fig.~\ref{fig:pcr_counts}. In particular we split the vaccinated user cohort into one that reports an infection prior to December 15, 2021 and one that reports PCR-tests after that date. We assume that for the former the infection was likely caused by B.1.617.2 (Delta) and for the latter it was caused by B.1.1.529 (Omicron). In analogy to Fig.~2 in the main manuscript and Fig.~\ref{fig:long_covid_delta} we compute average changes in RHR, step count and sleep duration for both cohorts, Fig.~\ref{fig:delta_omicron}. We further use a two-sided Welch t-test at a significance level of $\alpha=0.01$ to assess whether the respective averages have to be considered different. 

For all three vital signs we find similar qualitative temporal evolutions as well as magnitudes in the respective changes. Especially for resting RHR we find large a similarity in the weeks around a positive PCR-test when comparing breakthrough infections in the two respective cohorts, Fig.~\ref{fig:delta_omicron}A. Only from week two onwards does the return to baseline take a slightly different shape depending on the considered variant. However, at no point in time can the two averages be considered statistically different at a signigicance level of $\alpha=0.01$, indicating a likely similar imprint of an infection with either variant of concern on RHR.

Likewise, we find almost the same maximum reduction of ${\sim} 3,000$ steps per day in the week after a positive PCR-test regardless of the variant that caused the breakthrough infection, Fig.~\ref{fig:delta_omicron}B. Moreover, the respective averages only differ significantly during the week of the PCR-test and the second week after, Fig.~\ref{fig:delta_omicron}B. A visual inspection of Fig.~\ref{fig:delta_omicron}B suggests that the return to baseline takes slightly longer for a breakthrough infection with Delta which aligns with the common observation of a generally milder course of COVID-19 after an Omicron infection~\cite{nealon_omicron_2022, callaway2021bad}. 

Ultimate, we find again similar patterns across variants when considering average changes in sleep duration, Fig.~\ref{fig:delta_omicron}C. Average sleep duration increases by ${\sim 24}$ minutes per day for both breakthrough infections with B.1.617.2 or B.1.1.529. A visual inspection again suggests a somewhat slower return to pre-disease values for a Delta-infection compared to Omicron. This observation is underlined by a statistically significant difference in the expected sleep duration in the second week after the PCR-test, Fig.~\ref{fig:delta_omicron}.

Taken together, we conclude that it is reasonable to combine all recorded breakthrough infection into a single user cohort, since differences in vital changes between the two major variants B.1.617.2 and B.1.1.529 are small and hardly significant. Moreover, the analysis in Sec.~\ref{sec:omicron1} underlines that the results presented in the main manuscript do not change substantially on a qualitative level if only breakthrough infections before December 15, 2021 are considered in the vaccinated user cohort. Since B.1.617.2 is considered to be the variant of concern that causes severe courses of COVID-19 most often our analysis implies that the average vital changes in vaccinated users that suffer from such an infection are still lower than those of unvaccinated individuals that carried out an infection with B 1.1.7. (Alpha) or the wildtype.  

\section{Observed time differences between vaccinations and infection}

\begin{figure}[h!]
\centering
\includegraphics[width=.6\linewidth]{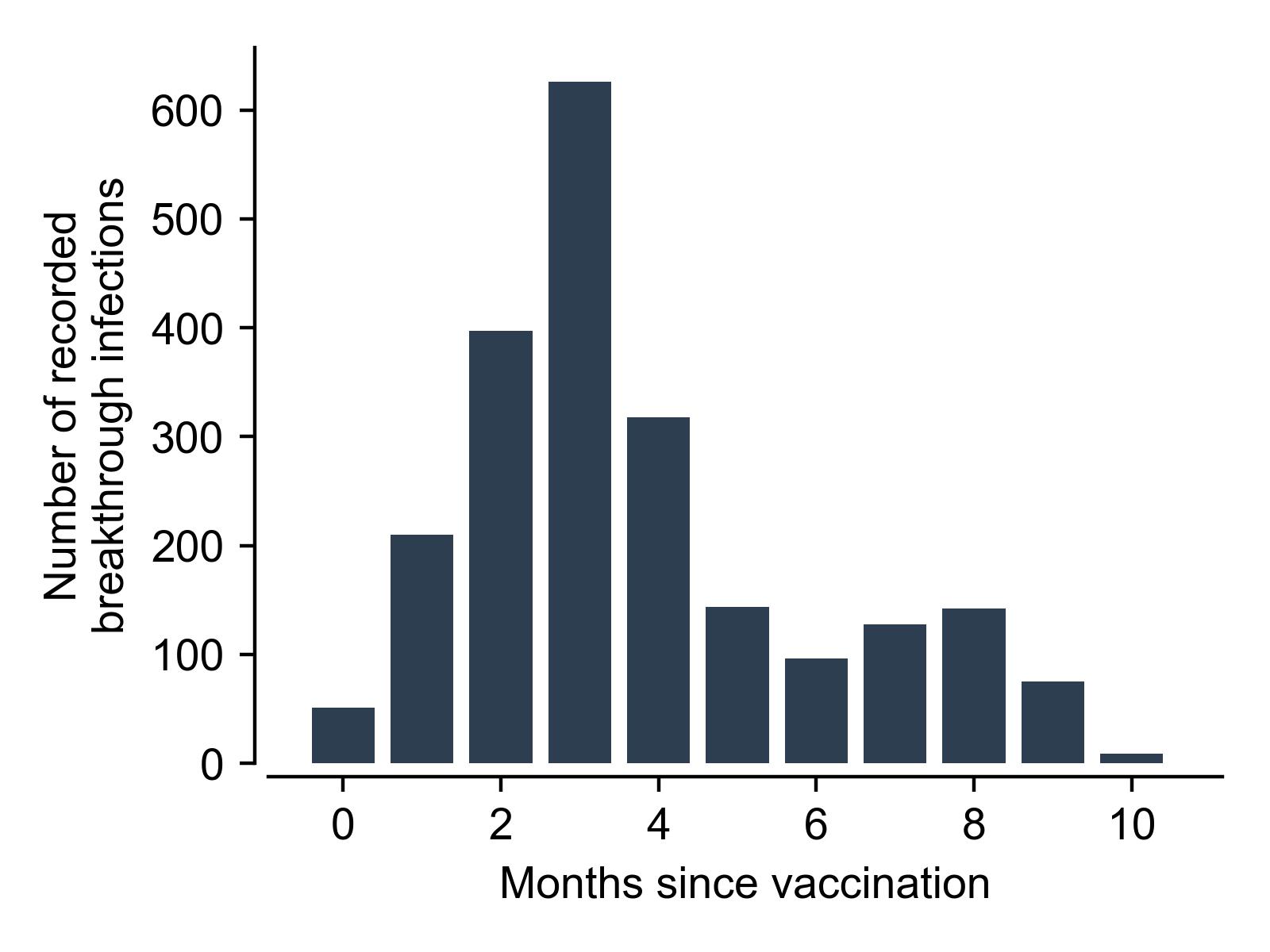}
\caption{Time difference in months between a confirmed infection with SARS-CoV-2 and receival of the last vaccination dose, i.e., the second or third dose depending on status. A time difference of $0$ months indicates that infection took place less than 30 days after receiving the last vaccination.}
\label{fig:time_diff}
\end{figure}

For all recorded breakthrough infections we compute the approximate time difference between receiving the last vaccination dose and the date of the PCR-test, Fig.~\ref{fig:time_diff}. Note that vaccination dates are only available with an accuracy of one month and PCR test dates only with an accuracy of one week. Hence, we cannot rule out that individuals contracted COVID-19 in the first two weeks after receiving the second vaccination dose, potentially misclassifying their case as a breakthrough infection since immunity might not have been achieved. However, only ${\sim} 2.3\%$ of all breakthrough cases are recorded in the month of the last vaccination dose. Hence, this effect can be considered negligible.  

\section{Representativeness of study cohort with respect to age}

\begin{figure}[h!]
\centering
\includegraphics[width=.6\linewidth]{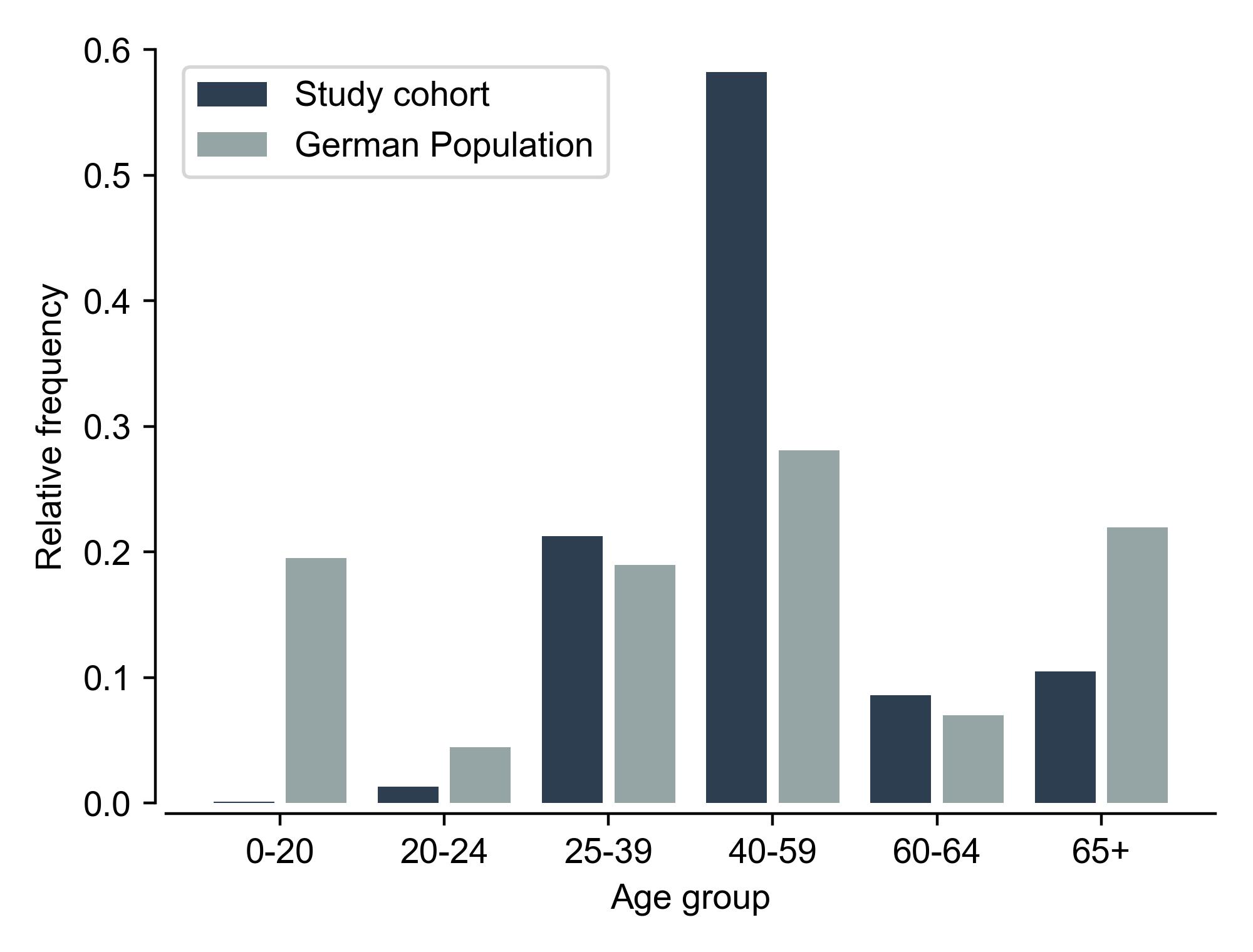}
\caption{Relative frequency of occurrences of age groups in our study cohort as well as the overall German population.}
\label{fig:age_groups}
\end{figure}

As already discussed in the main manuscript, our study cohort is not representative of the overall German population. However, a comparison with official census data from 2020~\footnote{obtained from \protect\url{https://www-genesis.destatis.de/genesis/online}} reveals that at least the frequencies of the commonly defined age groups 25-39 and 60-64 are well recaptured in our study cohort, Fig.~\ref{fig:age_groups}. We note a large over-representation of people aged 40-59 and consequentially an under-representation of elderly (aged 65 and more) and children/adolescents (aged 20 or younger). The latter group is by definition mostly excluded from our study since participation is only possible for citizens aged 16 or older. The elderly group is likely underrepresented due to a generally lower adoption of new technologies with older people.

%\bibliography{library}

\newpage

\section*{Supplementary Figures}

\begin{figure}[h!]
\centering
\includegraphics[width=\linewidth]{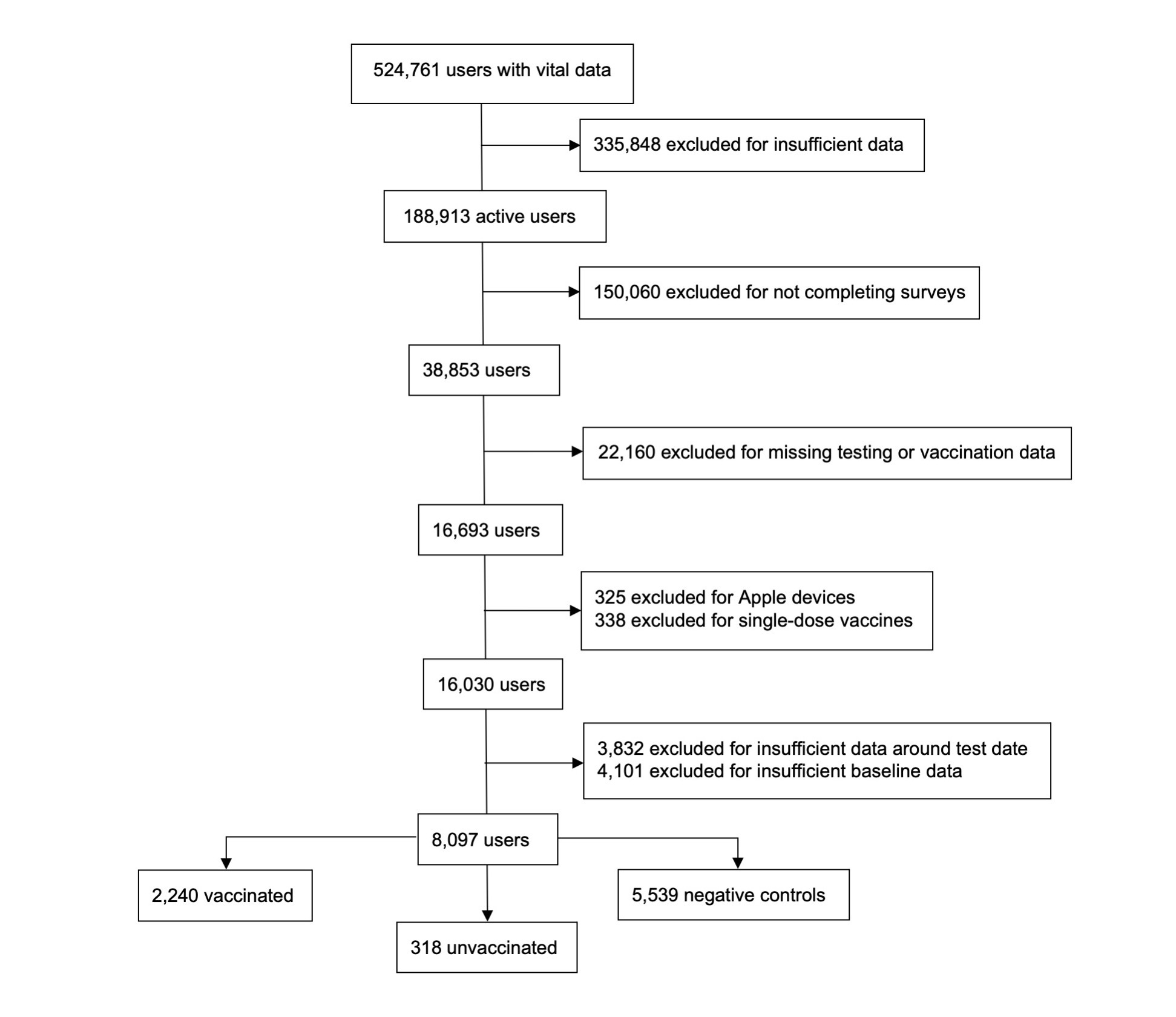}
\caption{Cohort diagram of the study group}
\label{fig:cohort}
\end{figure}

\end{document}